\documentclass[aps,pra,twocolumn,showkeys,nofootinbib,superscriptaddress]{revtex4}
\pdfoutput=1

\usepackage[utf8]{inputenc}
\usepackage{graphicx}
\usepackage{graphics}
\usepackage{epsfig}
\usepackage{lscape}
\usepackage{color}
\usepackage{multirow}
\usepackage{hyperref}
\usepackage{breqn}
\usepackage{physics}
\usepackage{subfiles}
\usepackage[singlelinecheck=false]{caption}
\usepackage{float}
\usepackage[export]{adjustbox}
\usepackage{amsfonts}
\usepackage{anyfontsize}

\newcommand{\Chemie}{
Department Chemie,
Johannes Gutenberg-Universit\"at, Fritz-Strassmann Weg 2, 55128 Mainz, Germany \\
}
\newcommand{\RUG}{
Van Swinderen Institute for Particle Physics and Gravity,
University of Groningen, Nijenborgh 4, 9747 Groningen, The Netherlands\\
}

\newcommand{\HIJ}{
Helmholtz-Institut Jena, Fr\"obelstieg 3, 07743 Jena, Germany\\
}
\newcommand{\GSI}{
GSI Helmholtzzentrum f\"ur Schwerionenforschung, 64291 Darmstadt, Germany\\
}
\newcommand{\US}{
Science Department, Chatham University, Pittsburgh, Pennsylvania 15232, USA\\
}
\newcommand{\FSU}{
Theoretisch-Physikalisches Institut, Friedrich-Schiller-Universit\"at Jena, D-07743 Jena, Germany\\
}
\begin{document}
\title{Laser Resonance Chromatography of $^{229}$Th$^{3+}$ in He: an \textit{ab} \textit{initio} investigation}
\author{Giorgio Visentin\footnote{Corresponding author: g.visentin@hi-jena.gsi.de}}
\affiliation{\HIJ} \affiliation{\GSI}
\author{Anastasia Borschevsky}
\affiliation{\RUG}
\author{Larry A. Viehland}
\affiliation{\US}
\author{Stephan Fritzsche}
\affiliation{\HIJ}
\affiliation{\GSI}
\affiliation{\FSU}
\author{Mustapha Laatiaoui}
\affiliation{\Chemie}
\date{\today}

\begin{abstract}
We propose a laser resonance chromatography (LRC) experiment on $^{229}$Th$^{3+}$, with the goal of detecting the ion's electronic ground $5f$ $^2$F$_{5/2}$ state and metastable $7s$ $^2$S$_{1/2}$ state by means of their ion mobilities. To this end, we first model the ion-neutral interaction potentials for the two electronic states with a relativistic Fock space coupled cluster method and complete basis-set extrapolation scheme. The interaction potentials are used to simulate the state-specific reduced ion mobilities in terms of the operating temperature and the external electric field. The ion mobilities differ by more than 7\% at 300 K and moderate field strengths; thus, separation of the $^{229}$Th$^{3+}$ metastable state lies within the reach of LRC experiments targeting optical probing and monitoring of the nuclear clock transition in this isotope. 
\end{abstract}
\maketitle

\section{Introduction}
\vspace{-5mm}
The atomic structure of $^{229}$Th has sparked increasing attention over the last decades \cite{Muller:2018, Fritzsche:2019}. This isotope exhibits a low-lying and relatively long-lived nuclear transition from the $^{229m}$Th isomer state at about 8.338(24) eV, which may be utilized in an optical nuclear clock of unprecedented precision and stability \cite{Thirolf:2024,Beeks:2021,Dzuba:2023,Seiferle:2017,Dzuba:2023b,Wense:2016,Muller:2018,Seiferle:2019,Peik:2003}. In particular, this clock could be used to investigate physics beyond the standard model, such as the space-time variation of the fundamental constants, violation of the Lorentz invariance and Einstein equivalence principle and search for dark matter fields \cite{Dzuba:2023, Dzuba:2023b,Muller:2018}. Although $^{229}$Th can be obtained in macroscopic quantities for spectroscopy purposes, the current challenge in this field is to precisely determine the nuclear clock transition so that it can be probed with laser radiation. Once the transition is found, advanced optical clock techniques could be used to stabilize the laser frequency and control the nuclear transition. However, investigation of the clock transition is challenged by competitive nuclear relaxation channels, which in turn depend on the electronic surrounding of the nucleus \cite{Seiferle:2017}. Internal conversion is the dominant nuclear relaxation channel in neutral Th and leads to the emission of one electron, that is favored by the relatively small ionization potential of the atom (6.3 eV) \cite{Seiferle:2017}. Ionic Th has thus been considered a more promising candidate as a nuclear clock compared to the neutral, as increasing charge states raise the ionization potential and make the internal-conversion channel energetically forbidden \cite{Seiferle:2017}. 
Th$^{3+}$ is nowadays considered the most promising candidate for a Th-based nuclear clock. This charge state possesses a relatively simple electronic structure that enables closed two- and three-level systems for laser-probing and cooling at the same time \cite{Block:2021}, and exhibits a laser-accessible 7s $^2$S$_{1/2}$ metastable state slightly above 22000 cm$^{-1}$ \cite{Safronova:2006} and with 1s lifetime \cite{Wense:2018} immune to field-induced frequency shifts \cite{Wense:2018,Block:2021}. Due to these features, Peik and Tamm \cite{Peik:2003} proposed a two-step laser-pumping scheme to populate the 7s $^2$S$_{1/2}$ electronic state: in the first step, the ground state is excited to the intermediate short-living 6d $^2$D$_{3/2}$ state, from which the metastable state is populated in the second step.

The next challenge to overcome in order to interrogate the nuclear clock transition of $^{229}$Th$^{3+}$ addresses the production and probing of the $7s$ $^2$S$_{1/2}$ metastable monitor electronic state. Here, we theoretically explore the application of Laser Resonance Chromatography (LRC) \cite{LRC,RomeroRomero:2022} for fast laser probing of a single isolated $^{229}$Th$^{3+}$ ion without the need for fluorescence detection. LRC specifically targets the investigation of heavy and superheavy ions, and characterizes them by means of their transport properties in buffer gases. 
A typical LRC experiment consists of a spectroscopic and chromatographic step. During the spectroscopic step, the ions are laser-excited to an electronic intermediate level and then to a dark long-living metastable state; the laser-pumping scheme for $^{229}$Th$^{3+}$ would include first the 1088 nm excitation from the electronic ground $5f$ $^2$F$_{5/2}$ state to the intermediate $6d$ $^2$D$_{3/2}$ state, followed by the 717 nm excitation to the metastable $7s$ $^2$S$_{1/2}$ state, according to the scheme proposed by Peik and Tamm \cite{Peik:2003}. In the chromatographic step, the ions in the ground and metastable states are driven through a drift tube filled with a buffer gas (usually an inert gas like He, Ne or Ar) and separated by means of their different transport properties (drift times). Variation of ions gas-phase transport properties with different electronic configurations, the so-called \textit{electronic}-\textit{state} \textit{chromatographic} (ESC) effect \cite{Visentin:2020}, is well established from both theoretical (e.g. \cite{Buchachenko:2022,Yousef:2007}) and experimental (e.g. \cite{Ibrahim:2008,Iceman:2007,Manard:2016}) investigations on the ion mobility of transition metal ions and lately lanthanide and actinide singly charged ions (e.g. \cite{Visentin:2020,Buchachenko:2014,Buchachenko:2019,Manard:2017,Ramanantoanina:2023}). LRC studies have been thus far restricted to singly charged cations drifting in helium; nonetheless, even multiply charged ions could be characterized by this technique, provided the buffer gas ionization potential exceeds the ion electron affinity \cite{Yamaguchi:2024}. The ionization potential of He (about 24.59 eV \cite{Kandula:2010}) is still larger than the ionization potential of Th$^{2+}$ (about 18.32 eV \cite{Wyart:1981}). Consequently, charge transfer from He to Th$^{3+}$ is unlikely, and the characterization of this triply charged ion should fall within reach of LRC.  Also, the different electron configurations associated with the ground $5f$ $^2$F$_{5/2}$ and $7s$ $^2$S$_{1/2}$ metastable states should ensure different and distinguishable ion mobilities. The challenge relies on the strength of the related ESC effect: the relative differences in the state-specific drift times and transport properties should lie above 7\% to enable efficient state separation \cite{LRC,Ramanantoanina:2023}. ESC effect can be maximized by proper selection of the operating temperature and external electric field that drive the ion through the drift tube \cite{LRC,RomeroRomero:2022}.

In this work, we simulated the chromatographic step of an LRC experiment on $^{229}$Th$^{3+}$ drifting in a helium gas, by, first, computation of accurate ion-atom interaction potentials for the two aforementioned electronic states and, second, simulation of the state-specific reduced (standard) ion mobilities over a wide range of gas temperatures and external electric field strengths, to determine the operating temperatures and field strengths that ensure state-separation. We did not consider interaction potentials and reduced ion mobilities associated with the ion in the low-lying 5f $^2$F$_{7/2}$, and intermediate 6d $^2$D$_{3/2}$ and 6d $^2$D$_{5/2}$ states, since these are too short-living to survive the spectroscopic step of the experiment and, thus, do not contribute to the subsequent chromatographic step. Our work is structured as follows: we detail the computational approach and methodology in Section II, while Section III is devoted to the discussion of the interaction potentials and ion mobilities computed according to Section II. Finally, the conclusions of our investigation are reported in Section IV. 

\vspace{-5mm}
\section{Methodology and Computational Details}
\vspace{-5mm}
\subsection{Ion-atom interaction potentials}
\vspace{-5mm}
The \textit{ab} \textit{initio} calculations on the Th$^{3+}$-He interaction potentials ($V(R)$) for the ground $5f$ $^2$F$_{5/2}$ and metastable $7s$ $^2$S$_{1/2}$ states were performed with the DIRAC23 \cite{DIRAC23} code. Relativistic effects were taken into account by means of the eXact-2-component (X2C) \cite{X2C} correction, and the nuclei were described by a finite-nucleus model \textit{via} the Gaussian charge distribution \cite{VisDya97}. The use of the X2C correction is justified by its extremely accurate performance with respect to the more computationally demanding four-component approach (see, for instance Refs. \cite{Dergachev:2023, Visentin:2021}).

For the basis set, we described the ion by means of the uncontracted Gaussian-type Dyall basis sets \cite{dyall2004,dyall2011} of double-augmented triple- (d-aug-v3z) and quadruple- (d-aug-v4z) zeta qualities, while the He atom was described by the augmented correlation-consistent polarized-valence triple- (aug-cc-pVTZ) and quadruple- (aug-cc-pVQZ) zeta basis sets of Woon and Dunning \cite{Woon1994}. This latter choice is motivated by the negligible relativistic effects featured by He.

The electronic structure was computed in two steps; first, Dirac-Hartree-Fock calculations were carried out on the closed-shell Th$^{4+}$. The resulting wavefunction was used as the reference wavefunction for the subsequent Fock space coupled cluster calculations with iterative single and double excitations (FSCCSD) \cite{Kaldor:1991,IH-FSCC} calculation on the (0,1) sector. To model the (0,0) sector, we correlated all the spinors from $4f$ up to those virtual spinors with energies below 30 $E_h$, with the inclusion of the $1s$ spinors of He. The model space to generate the (0,1) sector spanned the Th $5f$, $6d$ and $7s$ spinors, for a total of thirteen Kramers pairs. To these, one electron was added, whereas the remaining empty orbitals below 30 $E_h$ were treated as virtuals. By this approach, we computed the interaction potential in a range of interatomic distances from 1.0 up to 15 \AA. The interaction potentials were corrected for the basis-set superposition error by means of the procedure described by Boys and Bernardi \cite{Boys1970}. Eventually, we extrapolated the interaction potentials thus retrieved with the aforementioned finite basis set to the complete basis set (CBS) limit \cite{CBS}, by means of the Riemann $\zeta$-formula of Lesiuk and Jeziorski \cite{Lesiuk:2019}. In order to assess the accuracy of our calculations in the modelling of the electron correlation, we checked the extrapolated potentials for compliance with the ion-dipole asymptotic behavior in their long-range tail,
\begin{equation}\label{eq:LR}
V(R) = - (\alpha_{\text{He}}/2) R^{-4},
\end{equation}
where $\alpha$$_{He}$ stands for the static dipole polarizability of He, 1.38376 a.u. \cite{Schwerdtfeger:2019}. The largest deviation for the fitted He dipole polarizability was 0.02 a.u. (below 2\%) for all the interaction potentials.

\subsection{Ion mobilities}
\vspace{-5mm}
The mobility of an ion drifting in a buffer gas, $K$, is macroscopically defined as:
\begin{equation}\label{eq:IM}
\textbf{v}_d = K\textbf{E},
\end{equation}
where \textbf{v}$_d$ and \textbf{E} stand for the drift velocity of the ion and the external electric field, respectively. It is convenient to consider the reduced (standard) mobility, $K_0$:
\begin{equation}\label{eq:K0}
K_0 = n K/N_0,
\end{equation}
where $n$ refers to the number density of the buffer gas, whereas $N_0$ = 2.686705 $\times$ $10$$^{25}$ m$^{-3}$ is the Loschmidt number. In the framework of the two-temperature kinetic theory, the reduced mobility of an ion is defined in terms of the (buffer) gas temperature $T_0$ and the ratio of the electric field magnitude over the number density of the buffer gas, $E_0$/$n_0$ (i.e. the so-called reduced electric field)\cite{Viehland:2018}:
\begin{equation}\label{eq:2T}
K_0(E/n, T_0) = \left(\frac{2\pi}{\mu_0 k_B T_{eff}}\right)^{1/2} \frac{3q}{16N_0} \frac{1+\alpha_c(T_0)}{\bar{\Omega}^{(1,1)}(T_{eff})}.
\end{equation}
In \autoref{eq:2T}, $\mu_0$ is the reduced mass of the ion-atom system, $k_B$ is the Boltzmann constant, $T_{eff}$ is the effective temperature of the ion-atom system upon application of the electric field \cite{Viehland:2018}, $q$ is the ion charge and $\alpha_c$ is a corrective factor. The term $\bar{\Omega}^{(1,1)}(T_{eff})$ is called the collision integral and is defined as the temperature-average of the energy-dependent momentum-transfer cross sections \cite{Hirschfelder:1955}. Also, notice that in the low-field limit gas and effective temperatures coincide. 

We obtained the reduced ion mobilities from the solution of the Boltzmann equation for ion swarms by means of the Gram-Charlier approach \cite{Viehland:2018,Viehland:1994}. The computations were carried out in three steps; first, we calculated the momentum transfer cross sections as a function of the collision energy, with the help of the program PC \cite{Viehland:2010}. This quantity is obtained from the pre-computed ion-atom interaction potentials, which have to be as accurate as possible. To this end, we spliced our interaction potentials and shifted them in energy at the right internuclear distance grid endpoint to fit the ion-dipole interaction as retrieved from \autoref{eq:LR}; eventually, we used \autoref{eq:LR} to prolong our interaction potentials up to 40 \AA. The fractional accuracy on these calculations was set as low as 0.03\%. In the second step, we used the momentum-transfer cross sections to calculate the reduced ion mobilities in the low-field approximation, where the electric field strength is negligible and the reduced mobility depends only on the gas temperature (hereafter called the zero-field reduced ion mobility). The computations were performed by the program VARY \cite{Viehland:2012}, with a fractional accuracy set to 0.03\%. In the third step, we computed the reduced ion mobilities as a function of the reduced electric field and at four fixed temperatures, T $=$ 100, 200, 300 and 400 K. For these calculations we used the program GC \cite{Yousef:2007,Viehland:1994}, and set the fractional accuracy to 0.03\% for values of the reduced electric field below 60 Townsends (1 Td $=$ 10$^{-17}$ Vcm$^{2}$), 0.5\% between 60 and 64 Td, and 1\% at 100 Td. In the ion-mobility calculations we used the recommended value 229.031 7627(30) g/mole \cite{Wang:2012} as the atomic mass of $^{229}$Th.

Transport of an open-shell ion undergoes polarization and inelastic collision processes. In this work, we neglect collision-induced transitions between different terms of the ion, while we account for inelastic fine-structure transitions, which involve the projections of the ion total angular momentum $J$ along the interaction axis, i.e. $\Omega$. Indeed, this approximation holds for collisions occurring at moderate electric field strengths, such as those considered in this work, while strong-field regimes may imply non-negligible collision-induced transitions between different terms of the ion, such as collisional quenching of excited states. The scattering on multiple interaction potentials thus reduces to a single-channel event, under the assumption that two different extremes approximate the collision dynamics. 

The first extreme postulates statistical mixing of the projections $\Omega$ during the successive collision events. Thus, the drifting ion feels the carrier gas atoms through an effective averaged isotropic potential, $V_0$ \cite{Mason:2009}:
\begin{equation}\label{eq:ISO}
V_0 = \frac{2}{2J+1} \sum_{\Omega =1/2}^{J} V_{J\Omega},
\end{equation}
where $V_{J\Omega}$ refers to the interaction potential associated with a given projection $\Omega$ of the ion total angular momentum $J$. This extreme is called the \textit{isotropic} \textit{approximation} \cite{Visentin:2020, Buchachenko:2022, Buchachenko:2019,Buchachenko:2014,Aquilanti:1989}.

The second extreme assumes that the each collision event conserves $\Omega$. Each projection contributes to the collision event separately according to its statistical weight. The resulting reduced ion mobility is expressed as the average of the reduced ion mobilities for each projection $\Omega$ \cite{Ramanantoanina:2023,Visentin:2020}:
\begin{equation}\label{eq:ANISO}
K_0 = \frac{2}{2J+1} \sum_{\Omega =1/2}^{J} K_{0,J\Omega}.
\end{equation}
This extreme is called the \textit{anisotropic} \textit{approximation} \cite{Visentin:2020, Buchachenko:2022, Buchachenko:2019,Buchachenko:2014}.

\section{Results and Discussion}
\vspace{-5mm}
\subsection{Ion-atom interaction potentials}
\vspace{-5mm}
As far as to the authors' knowledge, no previous investigations exist on the interaction of Th$^{3+}$ with rare gases, and this restricts the direct assessment of our results. An indirect assessment can be performed by comparison of the $5f$ $^2$F$_{5/2}$ $\to$ $7s$ $^2$S$_{1/2}$ excitation energy with a literature analog. We thus estimated this excitation energy as the difference between the metastable and ground state total energies of the Th$^{3+}$-He dimer computed with the d-aug-v4z and aug-cc-pVQZ basis sets at 15 $\AA$, where the interaction with He is supposed to be negligible. We obtained an excitation energy equal to 21334.85 cm$^{-1}$, in good agreement with the 22229.0 cm$^{-1}$ analogous excitation energy computed by Safronova \textit{et} \textit{al.} \cite{Safronova:2006} with the all-order method with inclusion of single and double excitations.

\begin{figure}[H]
    \centering
    \includegraphics[width=\linewidth]{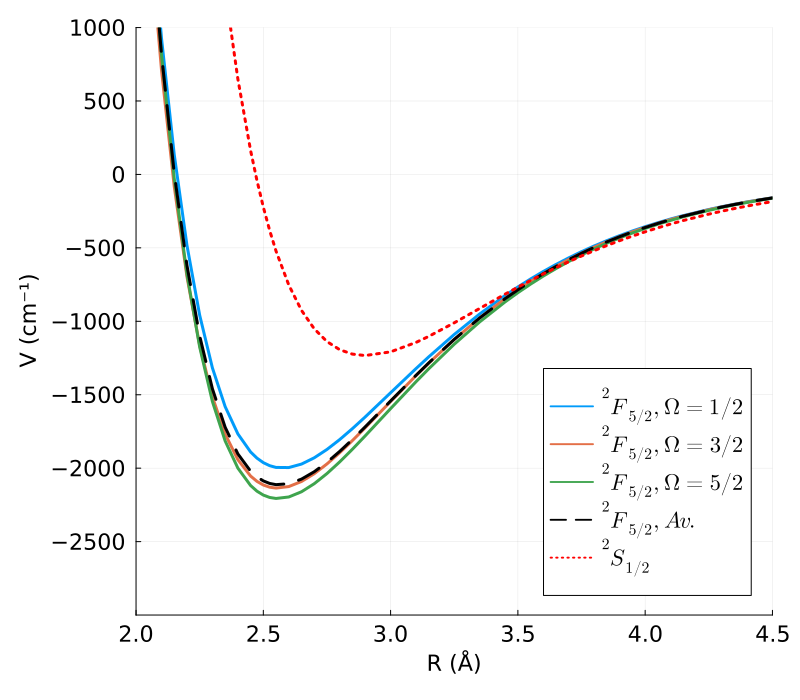}
    \caption{Graphical representations of the FSCCSD interaction potentials of the Th$^{3+}$-He system  calculated for the ground state $5f$ $^2$F$_{5/2}$ and the metastable state, $7s$ $^2$S$_{1/2}$. Counterpoise energy corrections are also included in the interaction potential. In particular, for the ground state, the interaction potentials associated with $\Omega$ = 1/2 (highest-lying full line), 3/2 and 5/2 (lowest-lying full line) are shown, along with the related average interaction potential (dashed line). For clarity the potentials are normalized to the same dissociation limit. }
    \label{PECs.png}
\end{figure}

In \autoref{PECs.png} we show the Th$^{3+}$-He interaction potentials for the ion's electronic ground state, $5f$ $^2$F$_{5/2}$ and the metastable state, $7s$ $^2$S$_{1/2}$. For the ground state, we also represent the isotropic potential obtained according to \autoref{eq:ISO}. In order to retrieve the spectroscopic parameters, i.e. the equilibrium distance $R_e$ and the dissociation energy $D_e$, we fitted the computed potentials from $R$ $=$ 1.0 to 4.5 $\AA$ to a Morse potential function \cite{Morse1929}:  

\begin{equation}\label{eq:Morse}
V(R) = D_e [e^{-2a(R-R_e)} - 2e^{-a(R-R_e)}].
\end{equation}
The use of the Morse potential function to fit the short-range region of heavy ion-rare gas interaction potentials is motivated by a previous work by Ramanantoanina \textit{et} \textit{al.} \cite{Ramanantoanina:2023}, where the spectroscopic parameters for the Lr$^+$-He interaction potentials for the ground and metastable states were retrieved with reasonable accuracy.
We list the spectroscopic and range ($a$) parameters in \autoref{table1}.

In the short-range region before 3.5 \AA, the interaction potential for the metastable $7s$ $^2S_{1/2}$ state is significantly more repulsive compared to the interaction potential for the ground $5f$ $^2F_{5/2}$ state. We ascribe this behavior to the ion's 7s orbital, that is oriented along the interaction axis as well as the He's 1s orbital, and thus enhances electron-electron Coulomb repulsion. For both ion states the interaction in the well region between 2.5 and 3 $\AA$  is strong, as the triply charged ion polarizes the 1s electrons of He; in particular, the interaction potential for the $\Omega$ = 5/2 projection of the ground state features a dissociation energy $D_e$ of roughly 2207 cm$^{-1}$, due to the strong ion-neutral induction interaction. This value is indeed in line with analogous spectroscopic parameters computed for the ground-state interaction potentials of multiply charged ions interacting with He, such as Mg$^{2+}$ (7574 cm$^{-1}$) and Ca$^{2+}$ (1240 cm$^{-1}$). However, when the ion is in its metastable state, the attractive dipole-induction term compete with the Coulomb repulsion originated from the singly occupied 7s orbital. This competition lowers by about 700 cm$^{-1}$ the dissociation energy for the ion in the metastable state with respect to the analog for the ion in the ground state.  For the ion ground state, the interaction is slightly anisotropic, with $\Omega$ = 5/2 and 3/2 giving rise to 10\% and 7\% deeper potential wells, respectively, compared to the $\Omega$ = 1/2 counterpart. We attribute this interaction anisotropy to a prolate zz-component of the permanent electric quadrupole moment associated with the Th$^{3+}$ $5f$ electron. This quantity is largest for $\Omega$ = 5/2, significant for $\Omega$ = 3/2,  but negligible for $\Omega$ = 1/2 (see \cite{Bellert:2002} for analogous discussions). 

Beyond 3.5 \AA, the interaction is dominated by dipole-induced and dispersion forces. As the former are isotropic, the interaction potentials associated with the ground state's $\Omega$ projections of the ion overlap, while the metastable-state analog falls slightly below the ground-state curves. We ascribe this behavior to the attractive dispersion forces that occur in the long-range region, and depend on the ion dynamic dipole polarizability. The $7s$ orbital in the ion metastable $7s$ $^2$S$_{1/2}$ state is more polarizable by the He 1s electrons compared to the more contracted 5f orbitals in the $5f$ $^2$F$_{5/2}$ state, and this explains the slightly attractive behavior of the former with respect to the latter. However, at longer internuclear distances dipole-induced terms prevail over dispersive ones and all the curves overlap.

\begin{table}[t]
\caption{Calculated range parameter $a$ (in \AA$^{-1}$), spectroscopic dissociation energy $D_e$ (in cm$^{-1}$) and equilibrium distance $R_e$ (in \AA) of the Th$^{3+}$-He interaction potentials in the ground 5f $^2$F$_{5/2}$ and metastable 7s $^2$S$_{1/2}$ states.} \label{table1}
\vspace{-5mm}
\begin{center}
\begin{tabular}{lllllllllllllllll}
\hline
\hline
$nl$ $^{2S+1}L_J$ &&&& \multicolumn{1}{c}{$\Omega$} &&&& \multicolumn{1}{c}{$a$} &&&& \multicolumn{1}{c}{$R_e$}   &&&& \multicolumn{1}{c}{$D_e$} \\
\hline
5f $^2$F$_{5/2}$  &&&& \multicolumn{1}{l}{1/2}      &&&& 1.665                   &&&& 2.576                       &&&& 1998.900                  \\
                  &&&& \multicolumn{1}{l}{3/2}      &&&& 1.681                   &&&& 2.558                       &&&& 2135.855                  \\
                  &&&& \multicolumn{1}{l}{5/2}      &&&& 1.692                   &&&& 2.557                       &&&& 2206.608                  \\
7s $^2$S$_{1/2}$  &&&& \multicolumn{1}{l}{1/2}      &&&& 1.598                   &&&& 2.904                       &&&& 1233.189                  \\
\hline
\hline
\end{tabular}
\end{center}
\end{table}

\subsection{Ion mobilities}
\begin{figure*}
\includegraphics[width=1.00\textwidth]{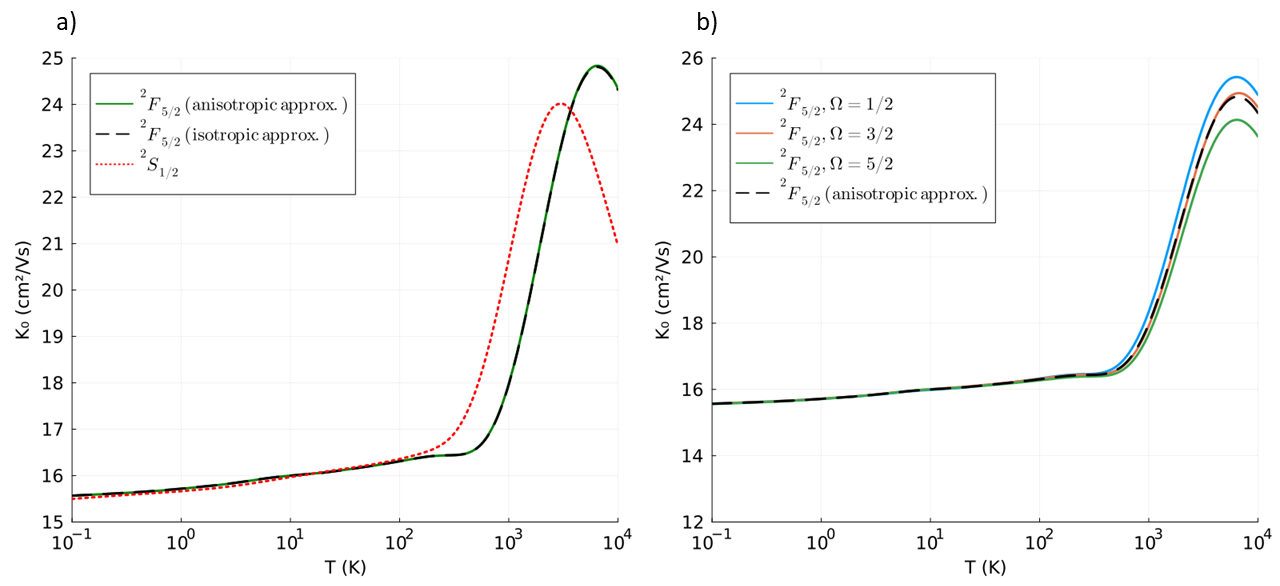}
\caption{a) Reduced zero-field mobilities of the $^{229}$Th$^{3+}$-He dimer in the ground $5f$ $^2$F$_{5/2}$ state (computed according to the isotropic and anisotropic approximations from \autoref{eq:ISO} and \autoref{eq:ANISO}, respectively) and metastable $7s$ $^2$S$_{1/2}$ state; b) zero-field mobilities associated with the $\Omega$ = 1/2 (highest-lying full line), 3/2, 5/2 (lowest-lying full line) projections of the ion's ground state compared with the average anisotropic mobility (dashed line).}
\label{IMT_NEW2.png}
\end{figure*}

\vspace{-5mm}
In \autoref{IMT_NEW2.png} we display the zero-field mobilities of $^{229}$Th$^{3+}$ in He computed over a wide range of gas temperatures. The figure consists of two panels; panel a) shows the mobilities for the ion ground state $5f$ $^2$F$_{5/2}$ computed according to both the isotropic and anisotropic approximations, and compare them to the mobility for the ion metastable state, $7s$ $^2$S$_{1/2}$; panel b) shows the mobilities associated with the three projections $\Omega$ of the ion ground state. The tabulated results are available in the Viehland database within the LXCat project \cite{LXCat}.

Let us first focus on panel a). Isotropic and anisotropic approximations yield results that differ by no more than 0.1\%. This similarity of the two extremes demonstrates the accuracy of our results \cite{Buchachenko:2014}. In the rest of this work, we will always refer to ion mobilities computed in the framework of the anisotropic approximations.

The gas-temperature range around 0.1 K coincides with the polarization limit at about 15.4 cm$^2$/Vs, where the mobility depends only on the dipole-induced contribution to the ion-atom interaction potential. The ion mobilities slowly grow with the gas temperature, until the mobility of the metastable-state ion starts increasing over the ground-state counterpart at around 100 K.  The mobility associated with the ion ground state exhibits a slight bump between 400 and 1000 K. This feature is absent in the ion mobility for the $7s$ $^2$S$_{1/2}$ state. The appearance of such concavities beyond the polarization limit had already been observed in the mobilities of singly charged coinage metal ions drifting in buffer gases heavier than Ne (see Ref.\cite{Yousef:2007}) and was ascribed to higher-order multipole-induced contributions to the ion-atom interaction. Panel b) shows that the depth of this concavity decreases with $\Omega$. Therefore, we explain this feature as due to the prolate zz-component of the permanent electric quadrupole moment associated with the 5f electron in Th$^{3+}$.  Both the state-specific ion mobilities feature a maximum at high gas temperatures, i.e., at 6400 K (24.83 cm$^2$/Vs) and 3100 K (24.15 cm$^2$/Vs) for the ground and metastable states, respectively. These high peak temperatures are due to the large dissociation energies of the related interaction potentials, as multiply charged ions strongly interact with rare gases (as shown in Ref. \cite{Viehland:2012} for the Mg$^{2+}$-He dimer).

\begin{figure}
\includegraphics[width=\linewidth, left]{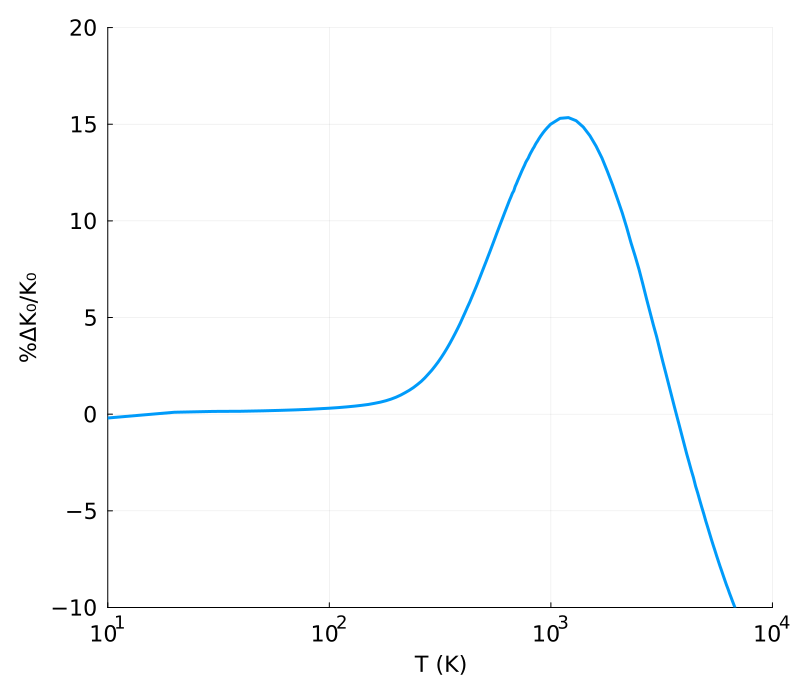}
\caption{Relative difference in the $^{229}$Th$^{3+}$-He zero-field reduced mobilities for the ion's ground $5f$ $^2$F$_{5/2}$ and metastable $7s$ $^2$S$_{1/2}$ states.}
\label{ESC.png}
\end{figure}

The ESC effect can be estimated from the difference of the state-specific ion mobilities \cite{Visentin:2020}:
\begin{equation}\label{eq:ESC}
\Delta K_0/K_0 = (K_0^* - K_0)/K_0,
\end{equation}
where we marked the mobility for the ion metastable state with an asterisk. In \autoref{ESC.png} we show the percentage difference in the state-specific zero-field ion mobilities. The mobility for the ion metastable state is higher than the ground-state analog by more than 7\% at 480 K, while reaches its maximum at about 1200 K, where $\Delta K_0$/$K_0$ is as large as 15\%. 480 K is already too high for a LRC experiment, as the operating temperatures should lie between 100 and 400 K. Therefore, we recommend the application of moderate reduced fields instead of higher temperatures. 

\begin{figure*}
\includegraphics[width=1.00\textwidth]{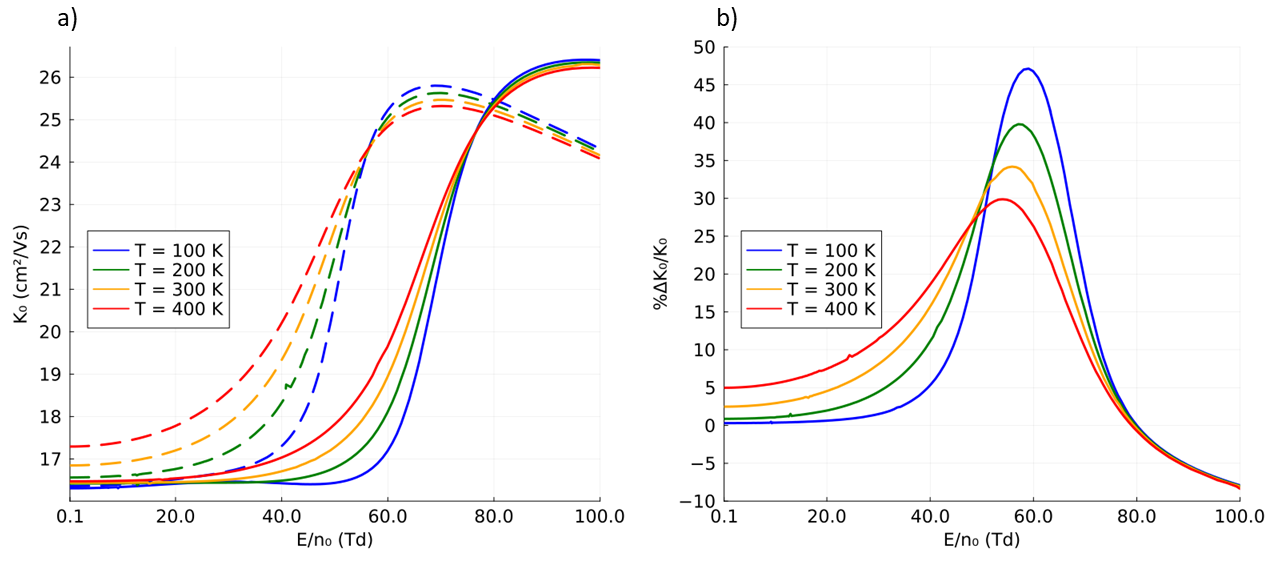}
\caption{a) Reduced mobilities of the $^{229}$Th$^{3+}$-He dimer in the ground $5f$ $^2$F$_{5/2}$ state (full lines) and metastable $7s$ $^2$S$_{1/2}$ state (dashed lines) as a function of the reduced electric field. The state-specific reduced ion mobilities are shown for four operating temperatures, from 100 K (lowest-lying full and dashed lines at low electric fields) to 400 K (highest-lying full and dashed lines at low electric fields), ; b) relative difference between the reduced zero-field mobilities of the ground and metastable states as a function of the reduced electric field. The reduced ion mobilities are shown for four operating temperatures, from 100 K (lowest-lying line at low electric fields) to 400 K (highest-lying line at low electric fields). }
\label{IMF_0-100_merged.png}
\end{figure*}

Let us now focus on \autoref{IMF_0-100_merged.png}. The figure shows the dependency of the $^{229}$Th$^{3+}$ mobility on the reduced electric field. In particular, panel a) shows the reduced mobilities of the ion ground $5f$ $^2$F$_{5/2}$ state and metastable $7s$ $^2$S$_{1/2}$ state, while panel b) represents the ESC effect estimated according to \autoref{eq:ESC}. Both quantities were computed at the four operating temperatures allowed by a typical LRC experiment.

We first explain panel a). The state-specific mobilities feature a similar behavior: at low field strengths, they grow with the temperature. The temperature-dependent reduced ion mobilities for the ground state cross at 70 Td, whereas the same behavior occurs at 45 Td for the metastable-state analogs.  At higher field values the reduced ion mobilities decouple from the temperature dependence, as the ion takes more energy from the electric field than from the thermal collisions with the atom \cite{Viehland:1975}. The crossings of the temperature-dependent mobilities for both states occur in the vicinity of the mobility maximum, where ion mobility is sensitive to the well of the interaction potential. The state-specific mobilities differ for three features: $i)$ the temperature dependence in the low-field limit, $ii)$ the concavity shortly before the crossing point and $iii)$ the position of the maximum. For the metastable state, the ion mobility is more sensitive to the temperature compared to the ground-state analog. In the low-field limit, the temperature is weakly sensitive to the reduced electric field, and we previously showed that at negligible fields the metastable state is more mobile than the ground state for gas temperatures between 100 and 400 K. Furthermore, the ground-state ion reduced mobility features a slower growth with the electric field compared to the metastable state counterpart, due to a concavity between 32 and 49 Td, that in turn is due to quadrupole-induced interactions. The position of the ion-mobility maximum relates to the ion-atom dissociation energy. 

An additional factor to take into account in order to enable efficient state separation by LRC is the strength of the applied electric field. Reduced electric fields should be as low as possible, in order to avoid collisional quenching of the metastable state \cite{LRC}. Optimal conditions for LRC should be found in a low-field range below 45 Td. Comparison of panels a) and b) of \autoref{IMF_0-100_merged.png} shows that in this range the reduced mobility is sensitive to the higher-order induction and dispersion contributions to the ion-atom interaction that arise shortly before the well region. The relative difference between the state-specific mobilities reaches 7\% at decreasing reduced field strengths as the operating temperature increases. In fact, at 100 K this difference is attained for reduced electric fields as large as 42 Td, while at 200, 300 and 400 K this 7\% relative difference is reached for 35, 28 and 18 Td, respectively. The best boundary conditions to ensure state separation and minimize collisional quenching of the metastable state relate to operating temperatures between 300 and 400 K and reduced electric fields between 18 and 28 Td. Temperatures close to room temperature are indeed recommendable from the experimental viewpoint, as the risk of thermal damage of the equipment is absent. However, further experimental investigations are needed in order to quantify the importance of collisional quenching processes with reduced fields between 20 and 30 Td. 

At higher fields the ESC effect reflects the decoupling of the ion mobility from the temperature dependence. Above 50 Td lower temperatures feature the largest differences in the state-specific mobilities. For instance, at 100 K and 60 Td the metastable state is 45\% more mobile than the ground state. 

\section{Summary and Conclusion}
\vspace{-5mm}
We proposed to use LRC to separate the electronic $7s$ $^2$S$_{1/2}$ state of $^{229}$Th$^{3+}$. This metastable state is currently investigated to monitor the nuclear clock transition in Th. In particular, we simulated the chromatographic step of the experiment, where metastable and ground electronic states are separated by means of their different interactions with a He buffer gas. We first computed the ion-atom interaction potentials for the ground $5f$ $^2$F$_{5/2}$ and metastable $7s$ $^2$S$_{1/2}$ states by accurate relativistic \textit{ab} \textit{initio} approaches. For both the electronic states of the ion the interaction with He is strong, albeit the potential energy curve for the metastable state is significantly shallower than its ground-state analog, due to Coulomb repulsion of the ion's $7s$ and the atom's $1s$ electrons at short internuclear distances. Strong dipole-induced forces and quadrupole-induced attraction featured by the $\Omega$ = 5/2 and 3/2 projections of the ground state make the interaction for this state strong and slightly anisotropic. This behavior and the different potential energy curves for the two electronic states encouraged us to simulate the state-specific ion mobilities in He, both in the framework of the low-field approximation and in the presence of growing electric fields. The state-specific reduced mobilities reflect the ion-atom interaction potentials, with the metastable-state ion being generally more mobile than the ground-state one over a wide temperature range. At 480 K, the metastable state is above 7\% more mobile than the ground state. This gas temperature is too high to be within reach of LRC; therefore we simulated the $^{229}$Th$^{3+}$ reduced mobility in He in the presence of increasing reduced electric fields and in a window of temperatures from 100 to 400 K. The state-specific reduced ion mobilities differ by more than 7\% in a temperature and reduced electric field ranges between 300 and 400 K, and 18 and  28 Td, respectively. These boundary conditions suit the operability of LRC and should enable effective state-separation, albeit further experimental investigations are needed to quantify collisional quenching rates of the metastable state.

Our investigation shows that LRC is a suitable technique to characterize the $7s$ $^2$S$_{1/2}$ electronic state that is needed for the interrogation of the $^{229}$Th nuclear clock transition, and, thus, may be integrated with experimental facilities devoted to the investigation of Th-based nuclear frequency standards.

Furthermore, the proposed FSCCSD-based computational approach is well suited for theoretical support of ion-mobility experiments, as ion-mobility is remarkably sensitive to dynamic electron correlation \cite{Buchachenko:2022}, which, in turn, is accurately described by methods based on the coupled cluster model. This characteristics makes the proposed computational approach scalable to the investigation of other heavy systems of interest for LRC experiments, that can be modelled by means of the FSCC level of theory. Such systems include, for instance, the ground and some low-lying excited states of Th$^{2+}$ (5f6d) and Ac$^{+}$ (7s$^2$), which can be modelled by means of the (0,2)-FSCC method, and No$^{+}$ (5f$^{14}$7s), which may be accurately modelled by means of the (0,1)-FSCC method.

\section{Acknowledgements}
\vspace{-5mm}
We gratefully acknowledge high-performance computing (HPC) support, time and infrastructure from the Ara cluster at the Scientific Computing and Data Science Centre of the Friedrich-Schiller University in Jena, Germany. M.L. acknowledges funding from the European Research Council (ERC under the European Union’s Horizon 2020 Research and Innovation Programme (Grant Agreement No. 819957). G.V. is also grateful to K.A. Peterson, A. I. Bondarev and Z.-W. Wu for suggestions and support. 
\bibliography{main}
\end{document}